\documentclass[twocolumn,showpacs,pra]{revtex4}

\usepackage{bm}

\usepackage{graphicx}
\usepackage{subfigure}

\newcommand{\ket}[1]{| #1 \rangle}

\newcommand{\cre}[1]{\hat{#1}^{\dag}}
\newcommand{\ann}[1]{\hat{#1}}

\begin{document}

\title{Quantum Phases of Ultra Cold Bosons in Incommensurate 1D Optical Lattices}

\author{Nir Bar-Gill}

\author{Rami Pugatch}

\author{Eitan Rowen}

\author{Nadav Katz}

\author{Nir Davidson}

\affiliation{Physics of Complex Systems, Weizmann Institute of Science, 76100
Rehovot, Israel}

\date{\today }


\begin{abstract}
We study theoretically a BEC loaded into an optical lattice in the tight-binding regime, with a second, weak incommensurate lattice acting as a perturbation. We find, using direct diagonalization of small systems and a large scale, number conserving Gutzwiller-based mean field approach, the phase diagram of the model.
We show that the superfluid, Mott-insulator and Bose-glass phases familiar from disordered systems appear here as well, with a localization transition occurring already in a 1D geometry, contrary to the well known Anderson localization. We complement these results with GPE simulations, and stress the fundamental difference between commensurate and incommensurate lattices. These results, together with calculations we have performed for realistic scenarios, suggest that this setup is well suited for available experimental realizations.
\end{abstract}

\maketitle


The possibility of observing quantum phase transitions by loading a Bose-Einstein Condensate (BEC) into an optical lattice \cite{greiner} has focused attention to disorder induced effects, based on the non-interacting Anderson phase transition \cite{anderson}. Although Anderson localization was observed (e.g. in microwave systems, see \cite{cheung}), the localization transition was not seen. Following the comprehensive treatment of the interacting case in \cite{fisher}, research specific to BECs in optical lattices in the presence of disorder has been done (e.g. \cite{lewenstein,batrouni2}). It was shown that the phase diagram of the system includes a compressible superfluid phase with a gapless excitation spectrum, a compressible (gapless) glass phase with no superfluid response, and an incompressible (gapped) insulator phase. 

Experimentally, attempts to create this form of random disorder were based on speckle patterns focused on the BEC \cite{inguscio}. The large correlation length of the disorder created in this method could not cleanly reproduce the localization transition, although reduced diffusion was measured \cite{inguscio2,aspect}. The possible observation of this localization transition is further complicated by the fact that it is present only in three dimensional systems (in the thermodynamic limit).

In this work we focus on a system consisting of a BEC in an optical lattice, to which a second optical lattice, with a wave vector incommensurate to the first, is added (\cite{lewenstein}). This second lattice is weak compared to the first, and acts as a perturbation similar in certain aspects to random disorder. Specifically, for non-interacting particles this model was considered in \cite{aubry}, where it was shown that it exhibits a localization transition as a function of the strength of the perturbation. 
Here we extend this model to the interacting case, and study the different phases as a function of both the interaction strength and the perturbation strength. A related setup, consisting of non-interacting atoms in two incommensurate lattices of equal strength, was studied in \cite{niu}. The problem of bosons in two-color optical lattices was also discussed in \cite{burnett}, in which commensurate lattices were studied. The fundamental difference between commensurate and incommensurate lattices will be discussed later. Recently, the model studied here was realized experimentally (\cite{inguscio3}).
We construct the phase diagram for incommensurate optical lattices, using an exact solution of the Bose-Hubbard Hamiltonian (\cite{jaksch}) for small systems. We employ the superfluid fraction and the determinant of the single-body density matrix ($\rho$) to distinguish between the phases \cite{pugatch}. It will be shown that for a mesoscopic system, the largest eigenvalue of $\rho$ distinguishes between the phases as well. We then perform large scale mean field calculations and study the system in a realistic experimental scenario. We show that the phase diagram exhibits the superfluid, Mott-insulator and Bose-glass phases, as well as effects of "disorder" induced order and delocalizing interactions. We also compare our system to the case of commensurate lattices, using both exact and mean field solutions of the Bose-Hubbard model, as well as GPE simulations. We then show that such incommensurability can be realized, and that our results are relevant for an actual experimental setup, taking into account limitations of finite size and harmonic confinement.

A system of bosons in a deep optical lattice in the tight binding regime is well described by the
Bose-Hubbard Hamiltonian, given by \cite{jaksch}
\begin{eqnarray}
\lefteqn{H = } \nonumber \\
&-&J \sum_{<i,j>} \cre{a_j} \ann{a_i} + \sum_{i} W_i \cre{a_i} \ann{a_i} + \sum_{i}
\frac{U}{2}\cre{a_i} \cre{a_i} \ann{a_i} \ann{a_i}, \label{eq:BH}
\end{eqnarray}
where $\cre{a_i}$ and $\ann{a_i}$ are spatial creation and annihilation operators obeying bosonic commutation relations. $J$ is the tunneling term between neighboring sites, $W_i$ is the on-site energy at site $i$, and $U$ is the on-site interaction energy. This model has been studied extensively \cite{bloch}, and is known to exhibit quantum phase transitions such as the superfluid - Mott insulator transition for $W_i=0$. 

Following Aubry and Andre \cite{aubry}, we first consider the homogeneous non-interacting limit, and introduce a second lattice with an incommensurate lattice vector as a perturbation. We choose the ratio of the wave vectors of the two lattices to be the golden ratio $\beta=\frac{1+\sqrt{5}}{2}$.
This perturbation alters the on-site energy term of Eq. (\ref{eq:BH}) to $W_i = \Delta cos (2 \pi \beta i)$, giving
\begin{equation}
H = -J \sum_{<i,j>} \cre{a_j} \ann{a_i} + \Delta \sum_{i} cos (2 \pi \beta i) \cre{a_i} \ann{a_i}. \label{eq:AA}
\end{equation}
Here $\Delta$ denotes the strength of the perturbation (or the depth of the incommensurate lattice). The addition of the second lattice alters the tunneling coefficient $J$ as well, but we neglect this effect (this will be discussed below). This single-particle Hamiltonian was shown \cite{aubry,aulbach} to have a localization transition in 1D, with the critical point at $\Delta/J = 2$, by transforming to the momentum basis. It is important to note the similarity between this Hamiltonian and the Anderson Hamiltonian, with the incommensurate lattice in the Aubry-Andre model playing the role of disorder in Anderson's model. However, an important difference between the models is that for disorder, a delocalization-localization transition occurs only in three dimensions or more. In the Aubry-Andre Hamiltonian such a transition can be seen already in one dimension.

We extend the Aubry-Andre model to include the effect of interactions, by adding an on-site repulsion term
\begin{eqnarray}
H &=& -J \sum_{<i,j>} \cre{a_j} \ann{a_i} + \Delta \sum_{i} cos (2 \pi \beta i) \cre{a_i} \ann{a_i} \nonumber \\
&+& \sum_{i}
\frac{U}{2}\cre{a_i} \cre{a_i} \ann{a_i} \ann{a_i}. \label{eq:AAU}
\end{eqnarray}
The Hamiltonian of eq. (\ref{eq:AAU}) can be solved by direct diagonalization under periodic boundary conditions \cite{burnett}. However, since the size of the matrix describing this many-body problem grows exponentially with the number of particles, we are limited to solving only small scale system, up to 9 particles in 9 sites. The properties of the system generally depend on the ratio of the number of particles to the number of sites, known as the filling factor. In this work we focus on the case of unit filling factor, in which the Mott-insulator phase appears. 

In studying this system we use the superfluid fraction \cite{fisher,leggett} and the determinant of the single-body density matrix ($\rho$) as the physical quantities which discriminate between the different phases \cite{pugatch}. The significance of $\rho$ in describing the single-body properties of the system and the condensate fraction is detailed in \cite{leggett}.
The quantity $det(\rho)$, which is base-independent, serves as a measure of the amount of fragmentation in the system. If the system is completely fragmented, such that the occupation of every single-body eigenfunction is non-zero, then $det(\rho) \neq 0$. Otherwise, the value of $det(\rho)$ is zero. It can be seen that this quantity is related to the energy gap, which has been used as an order parameter in similar problems. This is a result of the fact that for complete fragmentation, the energy gap is non-zero (with a value of $U$), and otherwise, since there are unoccupied states, is zero. However, $det(\rho)$ is an advantageous criterion, since it emerges directly as a property of the ground-state.

A widely used definition of the superfluid fraction is through the response of the system to transverse dragging of its boundaries. The dragging is assumed to be an adiabatic acceleration process, in which the boundaries reach a final, small velocity. This velocity can be described as a linear phase imprinted on the system, or a phase twist, and the superfluidity as the phase rigidity of the system. It can therefore be calculated \cite{burnett} for a phase twist $\theta$, with $N$ particles and $L$ sites through
\begin{equation}
f_s = \frac{L^2}{N J} \frac{E(\theta) - E(0)}{\theta^2}.
\end{equation}
In studying mesoscopic systems (as compared to taking the thermodynamic limit), $det(\rho)$ can be replaced by the largest eigenvalue of $\rho$, which is the condensate fraction, in distinguishing between the phases. In the thermodynamic limit in the presence of interactions, the condensate fraction is zero both in the Bose-glass phase and in the Mott-insulator phase, and therefore cannot distinguish between them. 

\begin{figure}[tbh]
\centering 
\includegraphics[width=1.0\linewidth,height=1.1\linewidth]{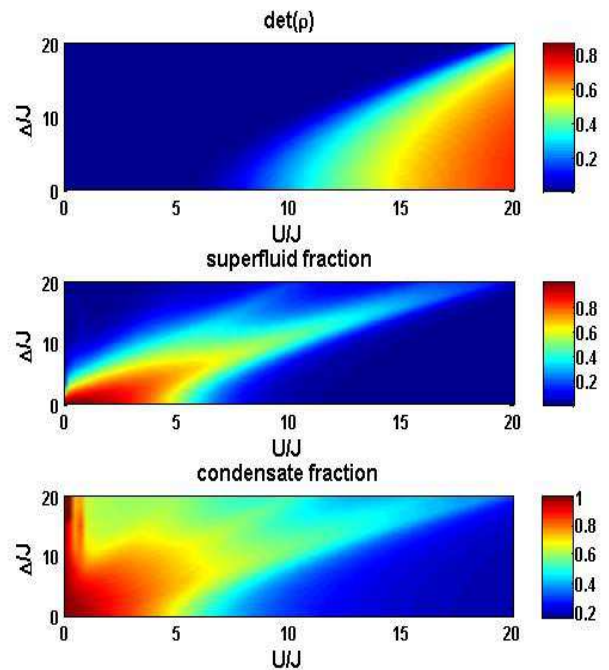}
\caption{$Det(\rho)$, superfluid fraction and condensate fraction as a function of interaction strength and perturbation strength, calculated by direct diagonalization of eq. (\ref{eq:AAU}) for 8 atoms in 8 sites. The first two quantities distinguish between the superfluid phase ($det(\rho) = 0, sf \neq 0$), the Bose-glass  phase ($det(\rho)=0,sf=0$) and the Mott-insulator phase ($det(\rho) \neq 0,sf=0$). The resulting phase diagram exhibits disorder induced order for the regime of large interactions and delocalizing interactions in the regime of strong disorder (see text). Since the system is of finite size, the condensate fraction can also be used instead of $det(\rho)$ to distinguish between the phases, as it is zero in the Mott-insulator phase, and non-zero otherwise.} 
\label{f1}
\end{figure}

In Fig. \ref{f1} we present the results of direct diagonalization of Hamiltonian (\ref{eq:AAU}) in one dimension. We show the values of $det(\rho)$, the superfluid fraction and the condensate fraction as functions of the interaction strength $U$ and the strength of the perturbation (the amplitude of the incommensurate lattice) $\Delta$, both in units of the tunneling energy $J$. Based on \cite{pugatch} we discriminate between the different phases of the system using $det(\rho)$ and the superfluid fraction. Thus we note that for small interaction strengths and small perturbations the system is in the superfluid phase, with $det(\rho) = 0$ and the superfluid fraction having a non-zero value. In this phase the mesoscopic condensate fraction (largest eigenvalue of $\rho$) is non-zero as well. By ``mesoscopic'' we mean a system of size $N$, in which to every value of $\Delta/J$ corresponds a finite value of $U/J$, such that interactions weaker than this value give a condensate fraction $cf >> 1/N$.
Keeping the strength of the perturbation small and increasing the strength of the interactions, the system passes into the Mott-insulator phase, where $det(\rho) \neq 0$ and the superfluid fraction is zero. Here the condensate fraction is also zero. For sufficiently strong perturbations, the system goes into the Bose-glass phase, in which the superfluid fraction is zero and $det(\rho)$ is also zero. In this case the condensate fraction is positive \cite{pugatch}. Specifically, for zero interaction strength we recover the Aubry-Andre transition from the superfluid phase to the localized phase as a function of the perturbation strength, with the transition point at $\Delta/J=2$. Checking this result for other system sizes, we find that the transition scales sharply to $\Delta/J=2$, as opposed to Anderson's transition, which scales to $\Delta/J=0$ for 1D \cite{anderson,pugatch}. At the other extreme, for $\Delta=0$, we recover the superfluid to Mott-insulator transition \cite{SubirQPT}. It is important to note that $det(\rho)$ and the superfluid fraction retain their distinction in the thermodynamic limit, and are therefore valid order parameters of the system. However, the condensate fraction scales to zero in the thermodynamic limit both for the Mott-insulator and for the interacting Bose-glass phase, and is therefore relevant only for mesoscopic systems. We have verified that indeed for larger systems $cf \rightarrow 0$ while $sf \rightarrow const$ (for the normalization $Tr(\rho) = N$).

The results of Fig. \ref{f1} show several similarities to the behavior of the Bose-Hubbard model with random quenched disorder. Most pronounced is the existence of the superfluid, Bose-glass and Mott-insulator phases and their relative position in phase space. Consequently, the effect of "disorder" induced order, i.e. the buildup of the superfluid fraction as a function of $\Delta$ in the strongly interacting regime, appears here. Also, the effect of delocalizing interactions, which is the buildup of a superfluid response as a function of $U$ in the regime of intermediate perturbation strengths, is observed in this model. Finally, fragmentation of the condensate in the Bose-glass phase for a mesoscopic system, which occurs for the case of disorder \cite{pugatch}, is also found for an incommensurate perturbation \cite{frag}. Fragmentation can be understood in the thermodynamic limit, by noting that above a certain strength of the perturbation, for every finite strength of interactions the lowest energy state will include a number of fragments which scales with the size of the system \cite{pugatch,cederbaum}. The transition from the Bose-glass phase to the Mott-insulator phase occurs when this fragmentation becomes complete fragmentation, i.e. one particle per site at unit filling.

Despite these similarities, the two forms of perturbations are inherently different, as indicated by the fact that in the Aubry-Andre model in 1D the transition from superfluid to Bose-glass occurs at a finite value of the perturbation, as opposed to the Anderson case, in which a transition exists only for three dimensions or more. Another interesting feature, which is apparent in Fig. \ref{f1}, are two stripes of a superfluid phase protruding from the central lobe of this phase. By studying the nature of the ground state of the system in this regime we find that these superfluid resonances are a consequence of energetic resonances between different many-body configurations. We have verified that such resonances exist for a variety of conditions, while their strength and position might vary.

Thus far we have based our discussion on results obtained for a small system through direct diagonalization of Hamiltonian (\ref{eq:AAU}). 
To verify that our main results are not a consequence of finite size artifacts, we also perform a large scale mean-field calculation based on the Gutzwiller ansatz \cite{rokshar}, adapted to the constraint of number conservation. In this method we separate the many-body wavefunction to single-site wavefunctions
\begin{equation}
| \psi \rangle = \prod_i \sum_n f_n^i |n \rangle,
\end{equation}
where $| n \rangle$ are particle number states, and $f_n^i$ are the corresponding amplitudes per site $i$. Minimizing the expectation value of the exact Hamiltonian (\ref{eq:AAU}) for $\ket{\psi}$ with respect to the $f^{i}_n$'s, gives us the estimated ground
state energy. In practice, in order to maintain a constant filling
factor, we need to add a chemical potential term $-\mu \sum_i n_i$
and to search for the minimizing $f^{i}_{n}$'s in the restricted
subspace where the total number of particles and the norms are
conserved. We use the equivalence between the Gutzwiller
ansatz and the mean-field approach
\cite{stoofopticallattices,SubirQPT} to get around the non-linear
minimization problem. This is done by diagonalizing the mean field Hamiltonian
at each lattice point,
and re-iterating this until a self consistent solution is reached,
again while conserving both number and norm. 



\begin{figure}[tbh]
\centering 
\includegraphics[width=1.0\linewidth]{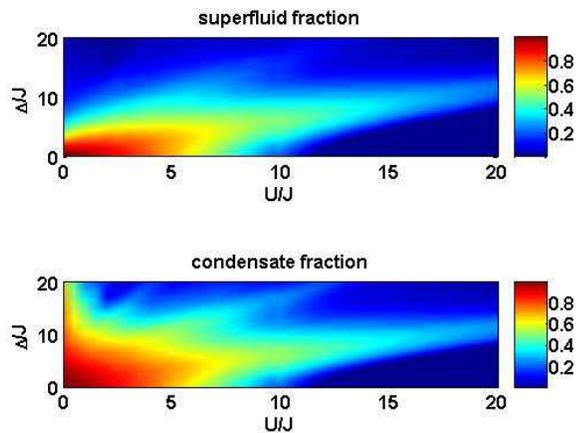}
\caption{Superfluid fraction and condensate fraction as a function of interaction strength and perturbation strength, calculated using the Gutzwiller based mean field approach for 60 atoms and sites. The phase diagram can be qualitatively reconstructed, despite the limitations of mean field for low dimensional systems, near phase transitions and for the small interactions regime. The condensate fraction is seen to qualitatively describe even the Bose glass phase for which it must approach zero at the thermodynamic limit, indicating the existence of a mesoscopic regime.} 
\label{f2}
\end{figure}
In Fig. \ref{f2} we show the results of the mean field calculation of the condensate and superfluid fractions as functions of both the interaction strength and the perturbation strength for a larger system with 60 sites and particles. It should be noted that due to the decoupled nature of the Gutzwiller wavefunction, this method becomes increasingly inaccurate as U/J lowers through the transition point and into the superfluid phase especially in 1D.
Moreover, it cannot reproduce the fragmentation of the condensate in the Bose-glass phase, which is seen in the exact diagonalization, and only sets an upper bound on the superfluid fraction (which is the average kinetic energy \cite{UpperBoundOnSF}). For the same reason, resulting from the fact that this calculation does not conserve $Tr(\rho)$, $det(\rho)$ cannot be calculated here, and the condensate fraction must be used as an indication of the different phases. Nevertheless, comparison between Figs. \ref{f1} and \ref{f2} reveals that the correct three phases of the system, as well as the superfluid phase resonances, are observed using mean field. The phase transitions themselves are smoothed out in the mean field
approach, and occur at different values compared to the analytical result (for $U=0$) and the exact solution (for $U>0$). Although the condensate fraction must approach zero for a localized ground-state at the thermodynamic limit, we find that it is non-zero for systems with as many as 120 particles and sites (it does however become smaller for larger systems with the same parameters).

\begin{figure}[tbh]
\centering 
\includegraphics[width=0.95\linewidth]{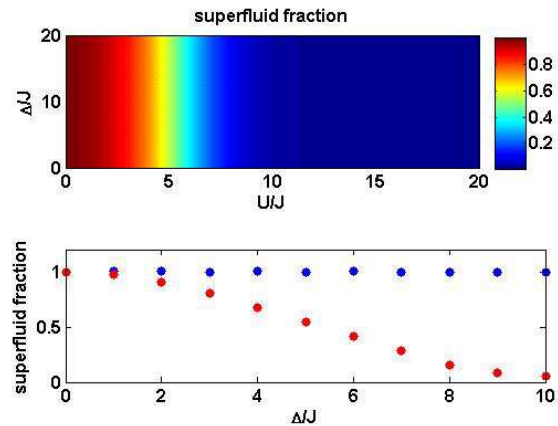}
\caption{Superfluid fraction as a function of interaction strength and perturbation strength, calculated using exact diagonalization (top), and the superfluid fraction as a function of $\Delta/J$ for $U/J=1$, calculated using the Gutzwiller based mean field approach (bottom), for commensurate (blue) and incommensurate (red) lattices. It can be seen that there is no dependence on $\Delta$, and the superfluid fraction only exhibits the superfluid to Mott-insulator transition. The mean field calculation shows that the superfluidity remains unity as $\Delta/J$ grows for commensurate lattices, while for incommensurate lattices it falls to zero.} 
\label{f3}
\end{figure}
The features of Hamiltonian (\ref{eq:AAU}) studied thus far are a direct consequence of the incommensurability of the two lattices. Similar properties were found in \cite{burnett} for commensurate lattices due to the finite size of the system examined therein. However, for thermodynamic and even mesoscopic systems, commensurate lattices exhibiting a periodic structure do not contain the phases found for incommensurate lattices. In Fig. 3a we plot the superfluid fraction as a function of $U/J$ and $\Delta/J$ found using an exact solution of Hamiltonian (\ref{eq:AAU}), with $\beta$ being a rational number, such that the two lattices are commensurate. We also plot the superfluid fraction as a function of $\Delta/J$ for $U/J=1$, found using our mean field approach, for both commensurate and incommensurate lattices. These calculations stress the difference between these two types of perturbations. Intuitively, incommensurate lattices create a periodic structure, for which the unit cell is determined by the ratio between the wavevectors of the two lattices. Since the combined potential is still perfectly periodic, such a system should not exhibit a Bose-glass phase, effectively eliminating the effect of the perturbation strength $\Delta$. In contrast, incommensurate lattices do not create a periodic structure, and therefore cannot be treated using Bloch theory. This fundamental difference gives rise to the localization phenomenon occurring as a function $\Delta$, and the appearance of the Bose-glass phase. In accordance with this intuitive view, it is clear from Fig. 3a that the values of the superfluid fraction along the $U/J$ axis are simply duplicated in the $\Delta/J$ axis, indicating that the perfect periodicity of the system renders the $\Delta/J$ axis redundant. This is in contrast to the superfluid fraction given in Figs. 1 and 2, for which a dramatic change occurs in the $\Delta/J$ direction, giving rise to localization and introducing the Bose-glass phase. 

\begin{figure}[tbh]
\centering 
\includegraphics[width=0.95\linewidth,height=0.5\linewidth]{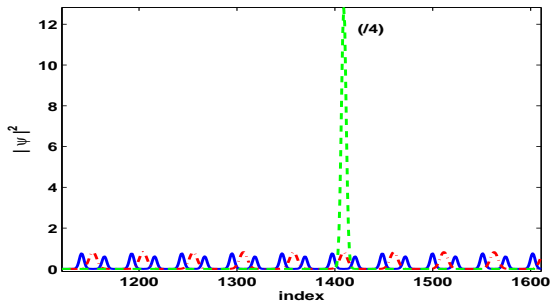}
\caption{Ground state real-space distribution for one lattice (blue solid), for a commensurate perturbation at $\Delta/J \sim 50$ (red dashed-dotted), and for an incommensurate perturbation at $\Delta/J \sim 50$ (green dashed, normalized to 1/4 to fit scale), calculated using a 1D GPE simulation. The localization is apparent for incommensurate lattices, as compared to commensurate lattices and to no perturbation.} 
\label{fgpe1d}
\end{figure}
We have also performed simulations of the Gross-Pitaevskii equation (GPE) in 1D, in the regime of weak interactions (for which the GPE approximation is valid - i.e. for this mesoscopic system the interactions are weak enough such that fragmentation in the Bose-glass phase is negligible), and assuming ``freezing out'' of the radial dynamics. Using GPE, the external potentials of the incommensurate lattices are used, and thus the values of $J$, $U$ and $\Delta$ are derived and not introduced as parameters. The difference between commensurate and incommensurate lattices is manifest using GPE simulations as well. In Fig. 4 we plot the ground state spatial distribution for the case of one lattice ($\Delta/J = 0$), and for the case of two lattices ($\Delta/J \sim 50$) for both the commensurate and the incommensurate cases. It is clear that the ground-state of commensurate lattices is delocalized, with a structure resulting from the periodic unit cell of the combined lattices, while the ground-state of incommensurate lattices exhibits a well-localized single peak.

Using our mean field method, we can further extend the Hamiltonian (\ref{eq:AAU}) to better describe realistic experiments. The Hamiltonian (\ref{eq:AAU}) can be approximately realized experimentally by loading a BEC into two optical lattices with incommensurate wave vectors, one of which is weak compared to the other. This incommensurate lattice may alter the tunneling energy $J$ as well as the on-site energy $U$, an effect we have neglected so far. By using the Gaussian tight-binding approximation \cite{jaksch,santos} we calculated the actual site-to-site tunneling coefficients and the on-site interaction terms as a function of $\Delta$. We have found that the variations in the interaction terms are completely negligible, and the tunneling energies vary by up to $~ 0.2\%$ for relatively weak perturbations ($\Delta \leq 30J$). In addition, GPE calculations for which the variations in $J$ and $U$ are taken into account inherently, show significant localization of the ground state as a function of the strength of the incommensurate lattice, which is apparent from both the spatial and momentum distributions (see Fig. 4).

Experimentally, for finite size systems, the incommensurability criterion states that the size of the unit cell of the combined lattices with momenta $k_1$ and $k_2$ must be larger than the size of the system. Moreover, the finite size $L$ of the system dictates a momentum accuracy $\Delta k \sim \hbar /L$ which the system can distinguish. 
It is therefore necessary to add to the criterion a requirement that for all the commensurate ratios $p/q$ such that $\left | p/q - k_1/k_2 \right| <\Delta k/k_2$, no $p/q$ will have a unit cell smaller than $L$.
We have verified that this combined criterion can be fulfilled for realistic conditions, by calculating using both Gutzwiller mean-field and GPE the superfluid and condensate fractions at $U/J=15$ for a system with 60 sites, and lattices with wavelengths of $784 nm$ and $1064 nm$. The resulting behavior is the same as for pure incommensurability (using the golden ratio).

\begin{figure}[tbh]
\centering 
\subfigure[]{
\includegraphics[width=0.48\linewidth,height=0.39\linewidth]{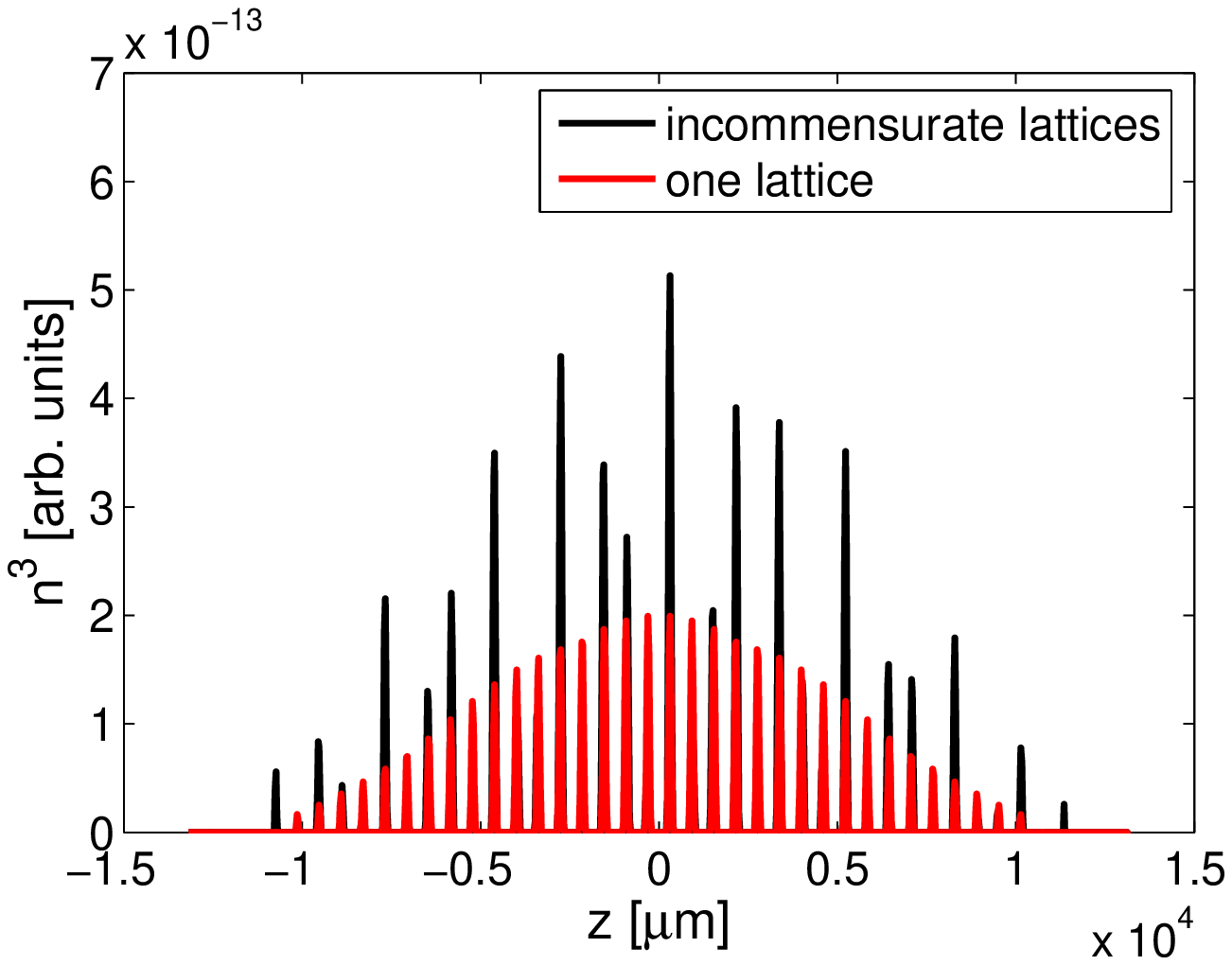}}
\subfigure[]{
\includegraphics[width=0.48\linewidth]{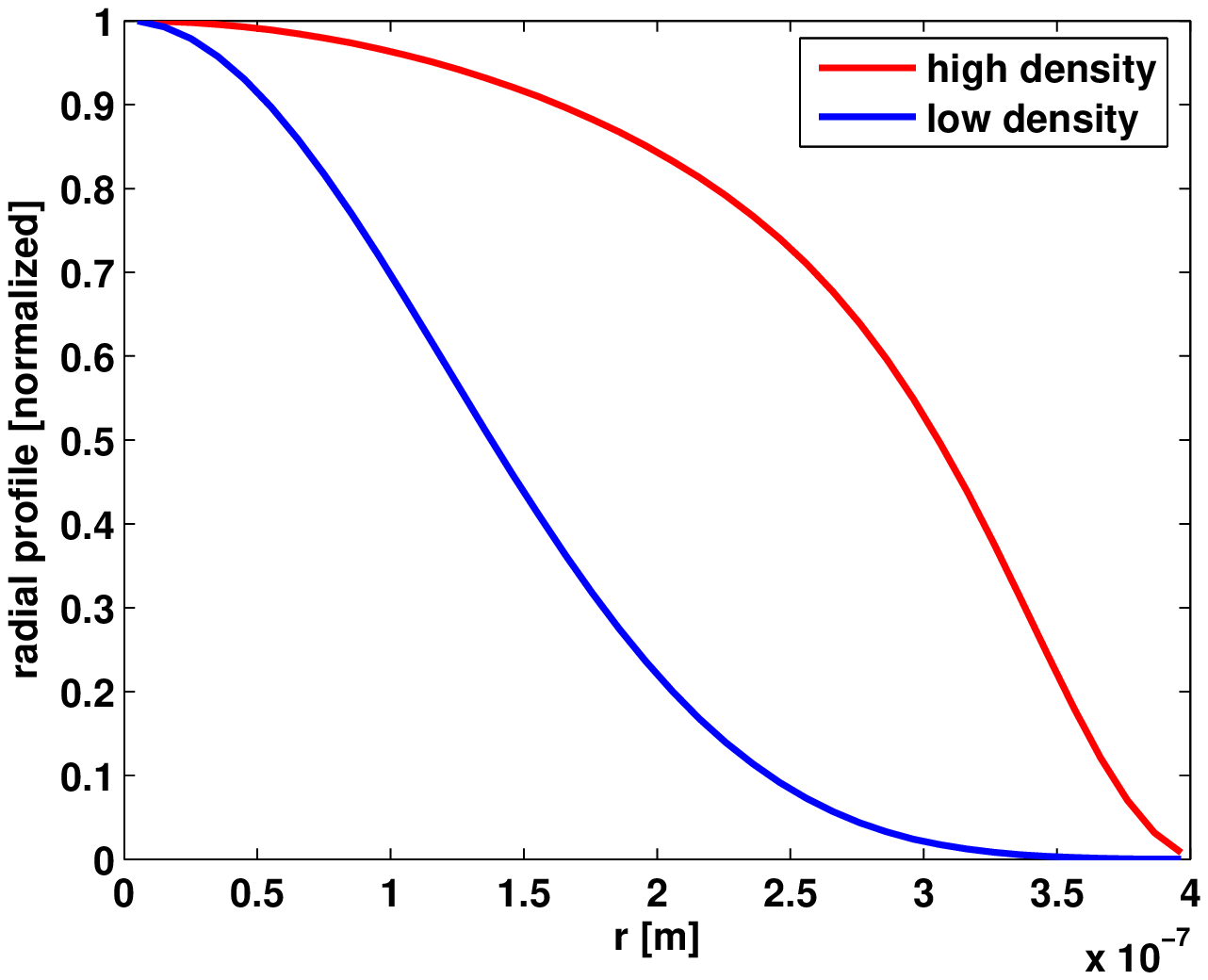}}
\caption{(a) Ground state spatial distribution calculated using a 3D GPE simulation, for $\omega_r=5kHz$ and $\mu=2kHz$, in the presence of one lattice (red) or incommensurate lattices (black). Compared to the truly 1D geometry, the localization here is less pronounced. (b) Radial density profiles of the ground state for a high density regime (red) and a low density regime (blue). While the low density regime retains a gaussian profile, at high densities the profile assumes the Thomas-Fermi shape.} 
\label{fgpe}
\end{figure}
An additional experimental consideration relates to the true dimensionality of the setup. We have thus far assumed a truly 1D system, which requires tight confinement in the transverse dimensions such that they are ``frozen out''. Naively, as long as the transversal trapping frequency $\omega_r$ is larger than the chemical potential $\mu$, the system can be treated as 1D. However, since the ground-state spatial distribution is irregular in the presence of the incommensurate lattice, this argument is no longer true. In Fig. 5a we plot the ground state wavefunction calculated using a 3D GPE simulation (see, e.g., \cite{katz}), assuming radial confinement of $\omega_r=5$ kHz and a chemical potential of $\mu=2$ kHz. Compared to the ground state in the presence of incommenusrate lattices calculated assuming a 1D geometry shown in Fig. 4, the spatial distribution here is clearly less localized. This suggests that despite the strong trapping in the transverse directions, the system does not retain its one-dimensional properties in the presence of the perturbation. This results in a less localized state, with a smoothed transition compared to the 1D case. To verify this we plot in Fig. 5b two radial profiles of the 3D ground-state, at longitudinal positions corresponding to high and low density centers. It can be seen that while the low density regime retains a gaussian profile in agreement with the ``transverse freezing'' assumption, the high density regime displays a markedly Thomas-Fermi like profile, which indicates significant transverse dynamics. We therefore note that standard magnetic or optical traps providing transverse trapping frequencies as high as a few kHz would not be able to create a truly 1D setup, and would not exhibit a well pronounced localization peak. In order to achieve one-dimensional behavior in the presence of localization, an array of 1D tubes created using a 2D optical lattice is needed.

\begin{figure}[tbh]
\centering 
\subfigure[]{
\includegraphics[width=0.48\linewidth]{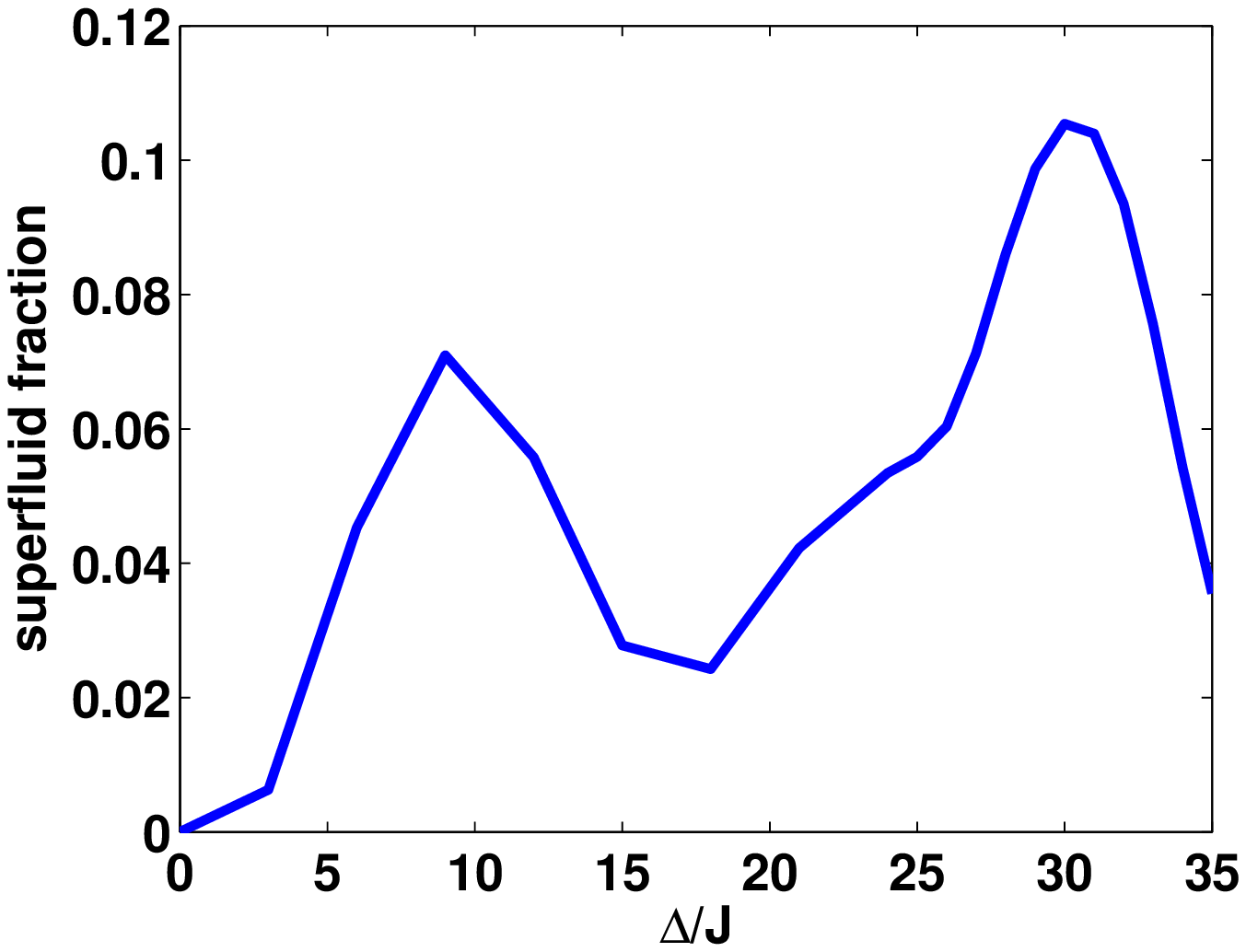}}
\subfigure[]{
\includegraphics[width=0.48\linewidth]{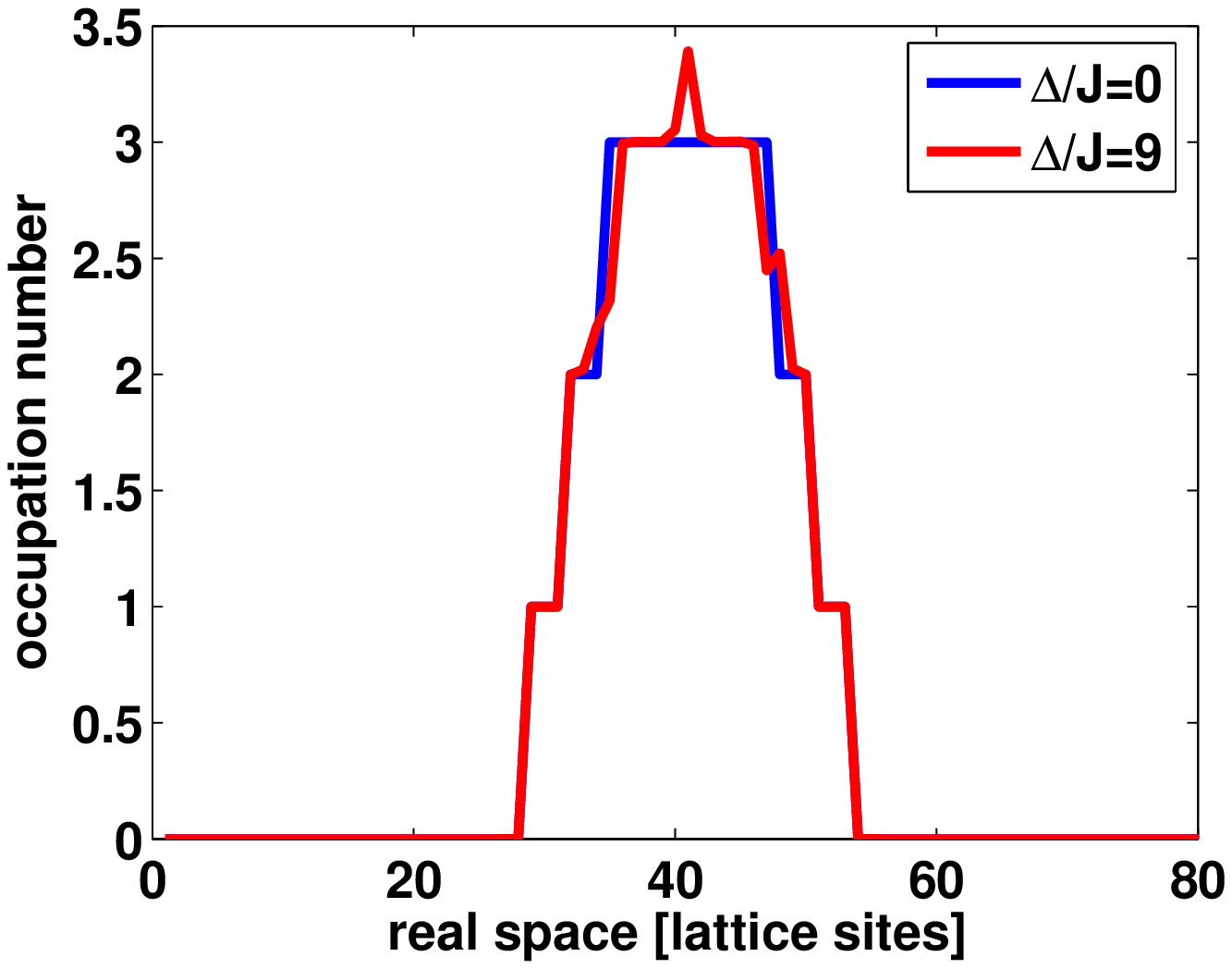}}
\caption{(a) Superfluid fraction as a function of perturbation strength for constant interaction strength, calculated using the Gutzwiller based mean field approach for 80 sites and 56 particles. $U/J=50$, with a confining harmonic potential, resulting in a smoothed and less pronounced transition from Mott-insulator to superfluid. Nevertheless, the superfluid fraction increases from approximately $0\%$ to $10\%$, which may be detected experimentally. The dependence of the superfluid fraction on $\Delta/J$ is sensitive to the specific form of the potential, but we have found that a response on the order of $10\%$ is a robust result. (b) The particle number distribution in space for $\Delta/J=0$ and $\Delta/J=9$ reveals the so-called "wedding-cake" structure \cite{jaksch,batrouni}, with Mott-insulator plateaus separated by superfluid regions. A small asymetric deformation of the structure can be observed as a result of the incommensurate perturbation.} 
\label{f4}
\end{figure}
Finally, we include the inhomogeneity of a realistic system, by adding to the on-site energy the effect of an harmonic confining potential. The effect of the confining potential for a one-lattice system in the Mott-insulator regime ($U/J = 50$) is to create a ``wedding cake'' structure in space (in terms of the occupation number per site) \cite{jaksch,batrouni}. By adding the incommensurate lattice as a perturbation, this structure is deformed, which results in an increase of the superfluid fraction of the system, approximately from $0\%$ to $10 \%$ (see Fig. \ref{f4}). The superfluid fraction can be measured by dragging the lattices adiabatically, and through time of flight images distinguishing the atoms that remain stationary. The localization transition can be identified experimentally by observing a sharp increase in the three-body loss caused by the high density at localized regions \cite{pugatch}.

In conclusion, we have studied a BEC in an optical lattice, with a second, incommensurate optical lattice acting as a weak perturbation. We have calculated the phase diagram of the system, using the superfluid fraction and the determinant of the single-body density matrix, and have recovered the three phases present in disordered models - superfluid, Mott-insulator and Bose-glass. For a mesoscopic system, we have shown that the condensate fraction can also distinguish between the phases (instead of $det(\rho)$), and indicates fragmentation in the Bose-glass phase which increases with interaction strength, up to full fragmentation in the Mott-insulator phase. However, in the thermodynamic limit, the condensate fraction vanishes in the Bose-glass phase for non-zero interactions, and therefore cannot act as an order parameter in that limit. Our solutions included both exact diagonalization of small systems and a new number conserving mean field method for large scale systems. We have complemented these calculations with 1D and 3D GPE simulations. These allowed us to study realistic experimental scenarios, including the effect of transverse trapping, varying tunneling energies and an harmonic confining potential. We have also used all three tools to demonstrate the fundamental difference between commensurate and incommensurate perturbations, as only the second exhibits localization. The similarities between this system in one dimension and disordered systems in three dimensions in terms of their phase diagrams, along with the fact that the perturbation we consider is created by introducing a second optical lattice, suggest the possibility of an experimental realization of our proposed setup. Such a setup is simpler than a 3D one, and overcomes the correlation length limitations of disorder introduced through speckle patterns.

This work was supported in part by the Israel Ministry of
Science, the Israel Science Foundation, and by the Minerva
Foundation.

\bibliography{bibBEC3}

\begin{thebibliography}{27}
\expandafter\ifx\csname natexlab\endcsname\relax\def\natexlab#1{#1}\fi
\expandafter\ifx\csname bibnamefont\endcsname\relax
  \def\bibnamefont#1{#1}\fi
\expandafter\ifx\csname bibfnamefont\endcsname\relax
  \def\bibfnamefont#1{#1}\fi
\expandafter\ifx\csname citenamefont\endcsname\relax
  \def\citenamefont#1{#1}\fi
\expandafter\ifx\csname url\endcsname\relax
  \def\url#1{\texttt{#1}}\fi
\expandafter\ifx\csname urlprefix\endcsname\relax\def\urlprefix{URL }\fi
\providecommand{\bibinfo}[2]{#2}
\providecommand{\eprint}[2][]{\url{#2}}

\bibitem[{\citenamefont{Greiner et~al.}(2002)\citenamefont{Greiner, Mandel,
  Esslinger, Hansch, and Bloch}}]{greiner}
\bibinfo{author}{\bibfnamefont{M.}~\bibnamefont{Greiner}},
  \bibinfo{author}{\bibfnamefont{O.}~\bibnamefont{Mandel}},
  \bibinfo{author}{\bibfnamefont{T.}~\bibnamefont{Esslinger}},
  \bibinfo{author}{\bibfnamefont{T.~W.} \bibnamefont{Hansch}},
  \bibnamefont{and} \bibinfo{author}{\bibfnamefont{I.}~\bibnamefont{Bloch}},
  \bibinfo{journal}{Nature} \textbf{\bibinfo{volume}{415}}, \bibinfo{pages}{39}
  (\bibinfo{year}{2002}).

\bibitem[{\citenamefont{Anderson}(1958)}]{anderson}
\bibinfo{author}{\bibfnamefont{P.~W.} \bibnamefont{Anderson}},
  \bibinfo{journal}{Phys. Rev.} \textbf{\bibinfo{volume}{109}},
  \bibinfo{pages}{1492} (\bibinfo{year}{1958}).

\bibitem[{\citenamefont{Zhang et~al.}(1998)\citenamefont{Zhang, Wong, Fung, Ho,
  Chan, Kan, Chan, and Cheung}}]{cheung}
\bibinfo{author}{\bibfnamefont{Z.~Q.} \bibnamefont{Zhang}},
  \bibinfo{author}{\bibfnamefont{C.~C.} \bibnamefont{Wong}},
  \bibinfo{author}{\bibfnamefont{K.~K.} \bibnamefont{Fung}},
  \bibinfo{author}{\bibfnamefont{Y.~L.} \bibnamefont{Ho}},
  \bibinfo{author}{\bibfnamefont{W.~L.} \bibnamefont{Chan}},
  \bibinfo{author}{\bibfnamefont{S.~C.} \bibnamefont{Kan}},
  \bibinfo{author}{\bibfnamefont{T.~L.} \bibnamefont{Chan}}, \bibnamefont{and}
  \bibinfo{author}{\bibfnamefont{N.}~\bibnamefont{Cheung}},
  \bibinfo{journal}{Phys. Rev. Lett.} \textbf{\bibinfo{volume}{81}},
  \bibinfo{pages}{5540} (\bibinfo{year}{1998}).

\bibitem[{\citenamefont{Fisher et~al.}(1989)\citenamefont{Fisher, Weichman,
  Grinstein, and Fisher}}]{fisher}
\bibinfo{author}{\bibfnamefont{M.~P.~A.} \bibnamefont{Fisher}},
  \bibinfo{author}{\bibfnamefont{P.~B.} \bibnamefont{Weichman}},
  \bibinfo{author}{\bibfnamefont{G.}~\bibnamefont{Grinstein}},
  \bibnamefont{and} \bibinfo{author}{\bibfnamefont{D.~S.}
  \bibnamefont{Fisher}}, \bibinfo{journal}{Phys. Rev. B}
  \textbf{\bibinfo{volume}{40}}, \bibinfo{pages}{546} (\bibinfo{year}{1989}).

\bibitem[{\citenamefont{Damski et~al.}(2003)\citenamefont{Damski, Zakrzewski,
  Santos, Zoller, and Lewenstein}}]{lewenstein}
\bibinfo{author}{\bibfnamefont{B.}~\bibnamefont{Damski}},
  \bibinfo{author}{\bibfnamefont{J.}~\bibnamefont{Zakrzewski}},
  \bibinfo{author}{\bibfnamefont{L.}~\bibnamefont{Santos}},
  \bibinfo{author}{\bibfnamefont{P.}~\bibnamefont{Zoller}}, \bibnamefont{and}
  \bibinfo{author}{\bibfnamefont{M.}~\bibnamefont{Lewenstein}},
  \bibinfo{journal}{Phys. Rev. Lett.} \textbf{\bibinfo{volume}{91}},
  \bibinfo{pages}{080403} (\bibinfo{year}{2003}).

\bibitem[{\citenamefont{Scalettar et~al.}(1991)\citenamefont{Scalettar,
  Batrouni, and Zimanyi}}]{batrouni2}
\bibinfo{author}{\bibfnamefont{R.~T.} \bibnamefont{Scalettar}},
  \bibinfo{author}{\bibfnamefont{G.~G.} \bibnamefont{Batrouni}},
  \bibnamefont{and} \bibinfo{author}{\bibfnamefont{G.~T.}
  \bibnamefont{Zimanyi}}, \bibinfo{journal}{Phys. Rev. Lett.}
  \textbf{\bibinfo{volume}{66}}, \bibinfo{pages}{3144} (\bibinfo{year}{1991}).

\bibitem[{\citenamefont{Lye et~al.}(2005)\citenamefont{Lye, Fallani, Modugno,
  Wiersma, Fort, and Inguscio}}]{inguscio}
\bibinfo{author}{\bibfnamefont{J.~E.} \bibnamefont{Lye}},
  \bibinfo{author}{\bibfnamefont{L.}~\bibnamefont{Fallani}},
  \bibinfo{author}{\bibfnamefont{M.}~\bibnamefont{Modugno}},
  \bibinfo{author}{\bibfnamefont{D.~S.} \bibnamefont{Wiersma}},
  \bibinfo{author}{\bibfnamefont{C.}~\bibnamefont{Fort}}, \bibnamefont{and}
  \bibinfo{author}{\bibfnamefont{M.}~\bibnamefont{Inguscio}},
  \bibinfo{journal}{Phys. Rev. Lett.} \textbf{\bibinfo{volume}{95}},
  \bibinfo{pages}{070401} (\bibinfo{year}{2005}).

\bibitem[{\citenamefont{Fort et~al.}(2005)\citenamefont{Fort, Fallani,
  Guarrera, Lye, Modugno, Wiersma, and Inguscio}}]{inguscio2}
\bibinfo{author}{\bibfnamefont{C.}~\bibnamefont{Fort}},
  \bibinfo{author}{\bibfnamefont{L.}~\bibnamefont{Fallani}},
  \bibinfo{author}{\bibfnamefont{V.}~\bibnamefont{Guarrera}},
  \bibinfo{author}{\bibfnamefont{J.~E.} \bibnamefont{Lye}},
  \bibinfo{author}{\bibfnamefont{M.}~\bibnamefont{Modugno}},
  \bibinfo{author}{\bibfnamefont{D.~S.} \bibnamefont{Wiersma}},
  \bibnamefont{and} \bibinfo{author}{\bibfnamefont{M.}~\bibnamefont{Inguscio}},
  \bibinfo{journal}{Phys. Rev. Lett.} \textbf{\bibinfo{volume}{95}},
  \bibinfo{pages}{170410} (\bibinfo{year}{2005}).

\bibitem[{\citenamefont{Clement et~al.}(2005)\citenamefont{Clement, Varon,
  Hugbart, Retter, Bouyer, Sanchez-Palencia, Gangardt, Shlyapnikov, and
  Aspect}}]{aspect}
\bibinfo{author}{\bibfnamefont{D.}~\bibnamefont{Clement}},
  \bibinfo{author}{\bibfnamefont{A.~F.} \bibnamefont{Varon}},
  \bibinfo{author}{\bibfnamefont{M.}~\bibnamefont{Hugbart}},
  \bibinfo{author}{\bibfnamefont{J.~A.} \bibnamefont{Retter}},
  \bibinfo{author}{\bibfnamefont{P.}~\bibnamefont{Bouyer}},
  \bibinfo{author}{\bibfnamefont{L.}~\bibnamefont{Sanchez-Palencia}},
  \bibinfo{author}{\bibfnamefont{D.~M.} \bibnamefont{Gangardt}},
  \bibinfo{author}{\bibfnamefont{G.~V.} \bibnamefont{Shlyapnikov}},
  \bibnamefont{and} \bibinfo{author}{\bibfnamefont{A.}~\bibnamefont{Aspect}},
  \bibinfo{journal}{Phys. Rev. Lett.} \textbf{\bibinfo{volume}{95}},
  \bibinfo{pages}{170409} (\bibinfo{year}{2005}).

\bibitem[{\citenamefont{Aubry and Andre}(1979)}]{aubry}
\bibinfo{author}{\bibfnamefont{S.}~\bibnamefont{Aubry}} \bibnamefont{and}
  \bibinfo{author}{\bibfnamefont{G.}~\bibnamefont{Andre}},
  \bibinfo{journal}{Coll. on Group Theoretical Methods in Physics, Israel}
  (\bibinfo{year}{1979}).

\bibitem[{\citenamefont{Diener et~al.}(2001)\citenamefont{Diener, Georgakis,
  Zhong, Raizen, and Niu}}]{niu}
\bibinfo{author}{\bibfnamefont{R.~B.} \bibnamefont{Diener}},
  \bibinfo{author}{\bibfnamefont{G.~A.} \bibnamefont{Georgakis}},
  \bibinfo{author}{\bibfnamefont{J.}~\bibnamefont{Zhong}},
  \bibinfo{author}{\bibfnamefont{M.}~\bibnamefont{Raizen}}, \bibnamefont{and}
  \bibinfo{author}{\bibfnamefont{Q.}~\bibnamefont{Niu}},
  \bibinfo{journal}{Phys. Rev. A} \textbf{\bibinfo{volume}{64}},
  \bibinfo{pages}{033416} (\bibinfo{year}{2001}).

\bibitem[{\citenamefont{Roth and Burnett}(2003)}]{burnett}
\bibinfo{author}{\bibfnamefont{R.}~\bibnamefont{Roth}} \bibnamefont{and}
  \bibinfo{author}{\bibfnamefont{K.}~\bibnamefont{Burnett}},
  \bibinfo{journal}{Phys. Rev. A} \textbf{\bibinfo{volume}{68}},
  \bibinfo{pages}{023604} (\bibinfo{year}{2003}).

\bibitem[{\citenamefont{Fallani et~al.}(2006)\citenamefont{Fallani, Lye,
  Guarrera, Fort, and Inguscio}}]{inguscio3}
\bibinfo{author}{\bibfnamefont{L.}~\bibnamefont{Fallani}},
  \bibinfo{author}{\bibfnamefont{J.~E.} \bibnamefont{Lye}},
  \bibinfo{author}{\bibfnamefont{V.}~\bibnamefont{Guarrera}},
  \bibinfo{author}{\bibfnamefont{C.}~\bibnamefont{Fort}}, \bibnamefont{and}
  \bibinfo{author}{\bibfnamefont{M.}~\bibnamefont{Inguscio}},
  \bibinfo{journal}{cond-mat} p. \bibinfo{pages}{0603655}
  (\bibinfo{year}{2006}).

\bibitem[{\citenamefont{Jaksch et~al.}(1998)\citenamefont{Jaksch, Bruder,
  Cirac, Gardiner, and Zoller}}]{jaksch}
\bibinfo{author}{\bibfnamefont{D.}~\bibnamefont{Jaksch}},
  \bibinfo{author}{\bibfnamefont{C.}~\bibnamefont{Bruder}},
  \bibinfo{author}{\bibfnamefont{J.~I.} \bibnamefont{Cirac}},
  \bibinfo{author}{\bibfnamefont{C.~W.} \bibnamefont{Gardiner}},
  \bibnamefont{and} \bibinfo{author}{\bibfnamefont{P.}~\bibnamefont{Zoller}},
  \bibinfo{journal}{Phys. Rev. Lett.} \textbf{\bibinfo{volume}{81}},
  \bibinfo{pages}{3108} (\bibinfo{year}{1998}).

\bibitem[{\citenamefont{Pugatch et~al.}(2006)\citenamefont{Pugatch, Bar-Gill,
  Katz, Rowen, and Davidson}}]{pugatch}
\bibinfo{author}{\bibfnamefont{R.}~\bibnamefont{Pugatch}},
  \bibinfo{author}{\bibfnamefont{N.}~\bibnamefont{Bar-Gill}},
  \bibinfo{author}{\bibfnamefont{N.}~\bibnamefont{Katz}},
  \bibinfo{author}{\bibfnamefont{E.}~\bibnamefont{Rowen}}, \bibnamefont{and}
  \bibinfo{author}{\bibfnamefont{N.}~\bibnamefont{Davidson}},
  \bibinfo{journal}{e-print cond-mat/0603571}  (\bibinfo{year}{2006}).

\bibitem[{\citenamefont{Mandel et~al.}(2003)\citenamefont{Mandel, Greiner,
  Widera, Rom, Hansch, and Bloch}}]{bloch}
\bibinfo{author}{\bibfnamefont{O.}~\bibnamefont{Mandel}},
  \bibinfo{author}{\bibfnamefont{M.}~\bibnamefont{Greiner}},
  \bibinfo{author}{\bibfnamefont{A.}~\bibnamefont{Widera}},
  \bibinfo{author}{\bibfnamefont{T.}~\bibnamefont{Rom}},
  \bibinfo{author}{\bibfnamefont{T.~W.} \bibnamefont{Hansch}},
  \bibnamefont{and} \bibinfo{author}{\bibfnamefont{I.}~\bibnamefont{Bloch}},
  \bibinfo{journal}{Phys. Rev. Lett.} \textbf{\bibinfo{volume}{91}},
  \bibinfo{pages}{010407} (\bibinfo{year}{2003}).

\bibitem[{\citenamefont{Aulbach et~al.}(2004)\citenamefont{Aulbach, Wobst,
  Ingold, Hanggi, and Varga}}]{aulbach}
\bibinfo{author}{\bibfnamefont{C.}~\bibnamefont{Aulbach}},
  \bibinfo{author}{\bibfnamefont{A.}~\bibnamefont{Wobst}},
  \bibinfo{author}{\bibfnamefont{G.-L.} \bibnamefont{Ingold}},
  \bibinfo{author}{\bibfnamefont{P.}~\bibnamefont{Hanggi}}, \bibnamefont{and}
  \bibinfo{author}{\bibfnamefont{I.}~\bibnamefont{Varga}},
  \bibinfo{journal}{New J. Phys.} \textbf{\bibinfo{volume}{6}},
  \bibinfo{pages}{70} (\bibinfo{year}{2004}).

\bibitem[{\citenamefont{Leggett}(2001)}]{leggett}
\bibinfo{author}{\bibfnamefont{A.~J.} \bibnamefont{Leggett}},
  \bibinfo{journal}{Rev. Mod. Phys} \textbf{\bibinfo{volume}{73}},
  \bibinfo{pages}{307} (\bibinfo{year}{2001}).

\bibitem[{\citenamefont{Sachdev}(1999)}]{SubirQPT}
\bibinfo{author}{\bibfnamefont{S.}~\bibnamefont{Sachdev}},
  \emph{\bibinfo{title}{Quantum phase transitions}}
  (\bibinfo{publisher}{Cambridge University Press}, \bibinfo{year}{1999}), ISBN
  \bibinfo{isbn}{0521582547}.

\bibitem[{fra()}]{frag}
\bibinfo{note}{Fragmentation is found by calculating the values of the other
  eigenvalues of the one-body density matrix. These are the second, third, etc.
  condensates, or fragments. In the Bose-glass phase in a mesoscopic system,
  the other condensates are non-zero, indicating fragmentation.}

\bibitem[{\citenamefont{Cederbaum and Streltsov}(2004)}]{cederbaum}
\bibinfo{author}{\bibfnamefont{L.~S.} \bibnamefont{Cederbaum}}
  \bibnamefont{and} \bibinfo{author}{\bibfnamefont{A.~I.}
  \bibnamefont{Streltsov}}, \bibinfo{journal}{Phys. Rev. A}
  \textbf{\bibinfo{volume}{70}}, \bibinfo{pages}{023610}
  (\bibinfo{year}{2004}).

\bibitem[{\citenamefont{Rokhsar and Kotliar}(1991)}]{rokshar}
\bibinfo{author}{\bibfnamefont{D.~S.} \bibnamefont{Rokhsar}} \bibnamefont{and}
  \bibinfo{author}{\bibfnamefont{B.~G.} \bibnamefont{Kotliar}},
  \bibinfo{journal}{Phys. Rev. B} \textbf{\bibinfo{volume}{44}},
  \bibinfo{pages}{10328} (\bibinfo{year}{1991}).

\bibitem[{\citenamefont{Dickerscheid et~al.}(2003)\citenamefont{Dickerscheid,
  van Oosten, Denteneer, and Stoof}}]{stoofopticallattices}
\bibinfo{author}{\bibfnamefont{D.~B.~M.} \bibnamefont{Dickerscheid}},
  \bibinfo{author}{\bibfnamefont{D.}~\bibnamefont{van Oosten}},
  \bibinfo{author}{\bibfnamefont{P.~J.~H.} \bibnamefont{Denteneer}},
  \bibnamefont{and} \bibinfo{author}{\bibfnamefont{H.~T.~C.}
  \bibnamefont{Stoof}}, \bibinfo{journal}{Phys. Rev. A}
  \textbf{\bibinfo{volume}{68}}, \bibinfo{pages}{043623}
  (\bibinfo{year}{2003}).

\bibitem[{\citenamefont{Paramekanti et~al.}(1998)\citenamefont{Paramekanti,
  Trivedi, and Randeria}}]{UpperBoundOnSF}
\bibinfo{author}{\bibfnamefont{A.}~\bibnamefont{Paramekanti}},
  \bibinfo{author}{\bibfnamefont{N.}~\bibnamefont{Trivedi}}, \bibnamefont{and}
  \bibinfo{author}{\bibfnamefont{M.}~\bibnamefont{Randeria}},
  \bibinfo{journal}{Phys. Rev. B.} \textbf{\bibinfo{volume}{57}},
  \bibinfo{pages}{11639} (\bibinfo{year}{1998}).

\bibitem[{\citenamefont{Gimperlein et~al.}(2005)\citenamefont{Gimperlein,
  Wessel, Schmiedmayer, and Santos}}]{santos}
\bibinfo{author}{\bibfnamefont{H.}~\bibnamefont{Gimperlein}},
  \bibinfo{author}{\bibfnamefont{S.}~\bibnamefont{Wessel}},
  \bibinfo{author}{\bibfnamefont{J.}~\bibnamefont{Schmiedmayer}},
  \bibnamefont{and} \bibinfo{author}{\bibfnamefont{L.}~\bibnamefont{Santos}},
  \bibinfo{journal}{Phys. Rev. Lett.} \textbf{\bibinfo{volume}{95}},
  \bibinfo{pages}{170401} (\bibinfo{year}{2005}).

\bibitem[{\citenamefont{Katz et~al.}(2005)\citenamefont{Katz, Rowen, Ozeri, and
  Davidson}}]{katz}
\bibinfo{author}{\bibfnamefont{N.}~\bibnamefont{Katz}},
  \bibinfo{author}{\bibfnamefont{E.}~\bibnamefont{Rowen}},
  \bibinfo{author}{\bibfnamefont{R.}~\bibnamefont{Ozeri}}, \bibnamefont{and}
  \bibinfo{author}{\bibfnamefont{N.}~\bibnamefont{Davidson}},
  \bibinfo{journal}{Phys. Rev. Lett.} \textbf{\bibinfo{volume}{95}}
  (\bibinfo{year}{2005}).

\bibitem[{\citenamefont{Batrouni et~al.}(2002)\citenamefont{Batrouni, Rousseau,
  Scalettar, Rigol, Muramatsu, Denteneer, and Troyer}}]{batrouni}
\bibinfo{author}{\bibfnamefont{G.~G.} \bibnamefont{Batrouni}},
  \bibinfo{author}{\bibfnamefont{V.}~\bibnamefont{Rousseau}},
  \bibinfo{author}{\bibfnamefont{R.~T.} \bibnamefont{Scalettar}},
  \bibinfo{author}{\bibfnamefont{M.}~\bibnamefont{Rigol}},
  \bibinfo{author}{\bibfnamefont{A.}~\bibnamefont{Muramatsu}},
  \bibinfo{author}{\bibfnamefont{P.~J.~H.} \bibnamefont{Denteneer}},
  \bibnamefont{and} \bibinfo{author}{\bibfnamefont{M.}~\bibnamefont{Troyer}},
  \bibinfo{journal}{Phys. Rev. Lett.} \textbf{\bibinfo{volume}{89}},
  \bibinfo{pages}{117203} (\bibinfo{year}{2002}).

\end{thebibliography}


\end{document}